
\input phyzzx
\PHYSREV
\newif\ifnonumbs
\nonumbsfalse
\advance\normalbaselineskip by -1pt
\ifnonumbs \nopubblock \nopagenumbers\fi
\date={May 1993}
\Pubnum={\caps UPR-566-T
 }

\def\to{\rightarrow}

\newdimen\instht
\newdimen\instwd
\newdimen\insrtpad
\insrtpad=3mm
\def\inst#1#2{\medbreak\begingroup\clubpenalty=10000
  \setbox5=\hbox{\vbox{#1}\hbox to \insrtpad{}}
  \instht=-\ht5
  \global\instht=\instht
  \setbox5=\vbox{\box5\vskip\instht}
  \def\par{\endgraf\endgroup\medbreak}
  \noindent\hang\hangafter=#2
  \instwd=\wd5
  \advance\instwd by \insrtpad
  \hangindent=\instwd
  \global\instwd=\instwd
  \hbox to 0pt{\hskip-\hangindent\box5\hfill}}
\outer\def\insrt#1#2{\inst{#1}{#2}}

\titlepage
\title{ Ordinary and Dilatonic Domain
Walls: Solutions and Induced Space-Times}%
\ifnonumbs\else\foot{A compilation of invited talks presented  at
 INFN Eloisatron Project 26th Workshop, ``From Superstrings to
Supergravity'',  Erice, Italy, December 5-12 1992 and SUSY
93 Conference, Boston, MA, March 29-April 2, 1993.}\fi
\frontpageskip=0.5\medskipamount plus 0.5 fil
\author{  Mirjam Cveti\v c}
\address{Department of Physics\break
University of Pennsylvania\break
Philadelphia, PA 19104--6396\break
}
\normalspace
\abstract{Recent developments in  unifying treatment  of
domain wall configurations and their global space-time structure is presented.
Domain walls  between vacua of non-equal cosmological constant
fall in three classes depending on the
value of their energy density  $\sigma$:
(i) extreme walls with $\sigma=\sigma_{ext}$   are planar, static  walls
corresponding to the supersymmetric configurations, (ii)
non-extreme walls  with $\sigma>\sigma_{ext}$  are expanding
bubbles with two insides, (iii) ultra-extreme walls with $\sigma<\sigma_{ext}$
 are bubbles of false
vacuum decay.
As a prototype exhibiting all  three types
of configurations  vacuum  walls between Minkowski
and anti-deSitter
vacua are discussed. Space-times associated with these walls exhibit
non-trivial causal
structure closely related to the one of the corresponding extreme and
non-extreme charged black holes, however, without singularities.
Recently discovered extreme dilatonic walls,  pertinent to
string theory, are also addressed.
They are static, planar  domain walls with metric in
the string frame being {\it flat} everywhere. Intriguing  similarities
between the  global space-time of  dilatonic walls and that of charged
dilatonic black holes are  pointed out.}
\endpage
\normalspace

\chapter{Introduction}

 \REF\VILENKIN {A. Vilenkin, Phys.\ Lett.\ {\bf 133B},  177 (1983).}

\REF\IS{J. Ipser and P. Sikivie, Phys.\ Rev.\  D {\bf 30},
712 (1984).}
\REF\VILI{For a review, see  A. Vilenkin, Phys. Rep. {\bf 121},
263 (1985).}

Domain walls\refmark{\VILENKIN ,\IS ,\VILI}  are the
most extended topological defects
\REF\BKT{ V. A. Berezin, V. A. Kuzmin, and I. I. Tkachev,
Phys.\ Lett.\ {\bf 120B}, 91 (1983).}
\REF\DESITTER{ H. Sato, Progr.\ Theor.\ Phys.\ {\bf 76}, 1250 (1986);
S. K. Blau, E. I. Guendelman, and A. H. Guth,
Phys.\ Rev.\ {\bf D35}, 1747 (1987);
V. A. Berezin, V. A. Kuzmin, and I. I. Tkachev,
Phys.\ Rev.\  {\bf D36}, 2919 (1987).}
\REF\CGRI{ M. Cveti\v c, S. Griffies and S.-J. Rey, Nucl.\ Phys.\
{\bf B381}, 301 (1992).}
arising in theories with isolated, in general
non-equal\refmark{\BKT ,\DESITTER ,\CGRI}
minima of the matter potential.
Domain walls can form as topological defects
in the early Universe in theories with isolated
minima of the matter potential.\Ref\Review{T. W. B.  Kibble, Phys. Rep.
 {\bf 67}, 183 (1980).}
 They also form  as boundaries of a true vacuum bubble created by a quantum
tunnelling process\Ref\Coleman{For a review, see S. Coleman, {\it Aspects of
Symmetry} (Cambridge Univ. Press, 1985).}
of the false vacuum decay, as well as
\REF\VIL{A. Vilenkin, Phys. Lett. {\bf 117B}, 25 (1982); Phys. Rev. {\bf D30},
 509 (1984).}
\REF\HH{J. B. Hartle and S. W. Hawking, Phys. Rev. {\bf D28}, 2960 (1983).}
\REF\LINDEI{A. Linde, Nuovo Cimento Lett. {\bf 39}, 401 (1984).}
 a boundary of the universes born from a
quantum tunnelling process from nothing (quantum cosmology).\refmark{\VIL ,
\HH , \LINDEI}

\REF\CGSII{M. Cveti\v c, S. Griffies and H. H. Soleng, {\it Non- and
Ultra-Extreme Domain Walls and Their Global Space-Times}, Univ. of
Pennsylvania preprint, UPR-546-T/Rev (March 1993).}
\REF\CGSIII{M. Cveti\v c, S. Griffies and H. H. Soleng, {\it Local and Global
Aspects of Domain Wall Space-Times}, Univ. of Pennsylvania Preprint, UPR-565-T
(May 1993).}
Recently, substantial progress\refmark{\CGSII ,
\CGSIII} has been made
in understanding the  space-time structure of  eternal vacuum
 domain wall configurations. The
work stemmed from an earlier discovery
\REF\CG{ M. Cveti\v c and
S. Griffies, Phys.\ Lett.\ {\bf 285B},  27 (1992).}
\REF\CGII{ For a review of supergravity walls,  see: M. Cveti\v c and
S. Griffies, in  {\em Proc.\ Int.\ Symp.\ on Black Holes,
Membranes and Wormholes\/}, The Woodlands, Texas, January 1992,
edited by S. Kalara and D. Nanopoulos  (World
Scientific, Singapore, in press).}
\REF\CQR{ Globally supersymmetric walls were studied by M. Cveti\v c,
 F. Quevedo and S.-J. Rey, Phys. Rev. Lett. {\bf 67},
1836 (1991) and
E. Abraham and P. Townsend, Nucl .Phys. {\bf B351}, 313 (1991).}
\REF\CDGS{ M. Cveti{\v{c}}, R. Davis, S. Griffies and H. H. Soleng,
Phys.\ Rev.\ Lett.\ {\bf 70}, 1191 (1993).}
\REF\Gibb{G. W. Gibbons,
Nucl.\ Phys. {\bf B394}, 3 (1993).}
 of supergravity  walls\refmark{\CGRI ,\CG ,\CGII,\CQR}\   and
a subsequent study  of
their global space-time structure.\refmark{\CDGS ,\Gibb}\
In particular, it was recognized\refmark{\CGSII ,\CGSIII} that
the extended nature of these
defects imposes strong constraints on the topology of the
wall and the nature of its
 space-time.
It turns out that  the   walls provide a fertile ground to study
globally non-trivial space-times. These space-times were
 found\refmark{\CDGS ,\CGSII} to be
 closely related to the ones of certain black holes, however,
without singularities.

``Ordinary'' domain walls  between vacua of non-equal cosmological constant
fall into three classes:\refmark\CGSII\
 (i) extreme  (supersymmetric)  static, planar  domain
 walls,\refmark{\CGRI, \CG, \CGII ,\CDGS,\Gibb}
 (ii) non-extreme domain
walls (expanding bubbles with an inertial observer  inside the bubble
for  each side of the wall)\refmark{\CGSII ,\IS}\ and  (iii)
ultra-extreme  walls (expanding
bubbles\refmark{\BKT ,\DESITTER}\ of false vacuum
\REF{\CD} {S. Coleman and F. De~Luccia, Phys.\ Rev.\  {\bf D21},
3305 (1980). }
decay\refmark\CD).  The energy density $\sigma_{non}^{ultra}$
of the non- [or ultra-]extreme walls
is bound from below [or above]
 by the one $\sigma_{ext}$
of the extreme ones.
Walls  are thus an example\refmark\CGSII\  of
configurations  for which   supersymmetry provides
a   lower   bound for the energy of stable wall
configurations.\Ref\KalGib{ In the black hole context, see
G. W. Gibbons in: {\em Supersymmetry,
Supergravity and Related Topics,\/} edited by F. del Aguila et al.\
(World Scientific, Singapore 1985), p. 147;
R. E. Kallosh, A. D. Linde, T. M. Ort{\' {\i}}n, A. W. Peet,
and A. van Proeyen, Phys.\ Rev.\   {\bf D46}, 5278 (1992).}
A prototype which exhibits all three types of walls are   walls
 between anti-deSitter and Minkowski vacuum.
The space-times induced by  such  walls are non-singular
with non-trivial global structure
and  horizons closely related to the
ones of certain black holes: on the
anti-deSitter [or Minkowski] side of the wall the induced
non-singular space-time is closely related to the ones
 of  the
Reissner-Nordstr\" om  [or Schwarzschild] black holes.\refmark{\CGSII
,\CGSIII}\
In the domain wall case the role of the mass ($M$) and
the charge ($Q$) of the black hole is played by the
energy density ($\sigma$)  of the
wall  and the cosmological constant ($\Lambda$)
outside the wall, respectively.

Recently discovered dilatonic domain walls,\Ref\CVETIC{M. Cveti\v c, {\it Flat
world of Dilatonic Domain Walls}, Univ. of Pennsylvania preprint, UPR-560-T
(April 1993).}\ on the other hand, are
solutions specific to superstring
theory. In such a domain wall
background along with the  matter fields
and  metric also
 the dilaton field\Ref\HL{ For an attempt  to study domain walls within the
Jordan-Brans-Dicke theory
see H.-S. La, CTP-TAMU-52/92 hepph/9212041, CTP-TAMU-78/92
hepth/9207202.}
 changes its value.  Dilatonic domain
walls are of particular interest because they correspond to
configurations which  interpolate between isolated
superstring vacua and may thus
shed light on the nature and connectedness  of superstring vacua.
\REF\CBH{G .Gibbons, Nucl.  Phys. {\bf B207}, 337 (1992);
G. Gibbons and K.
Maeda, Nucl. Phys. {\bf B298}, 741 (1988);
D. Garfinkle, G. Horowitz and A.
Strominger, Phys. Rev. {\bf D43}, 3140 (1991), {\bf D45},
3888 (1992){\bf E}.}
\REF\HOR{ For a review,  see G. Horowitz, {\it The
Dark Side of String Theory: Black Holes and Black Strings}, USCB-92-32
(October 1992) and references therein.}

Dilatonic   domain walls  are a generalization
of the ordinary   domain
walls in an analogous way as   dilatonic charged black
holes\refmark{ \CBH ,\HOR} are a generalization of
``ordinary'' black holes.
  The intriguing similarity between
the  space-time of the walls and   the
 one of the corresponding black-holes  reappears  in
the case of dilatonic walls as well.

The rest of the paper is organized in the following way.
In  Chapter 2  the space-time of
ordinary domain walls is discussed. In Chapter 3 dilatonic walls are addressed.
Brief conclusions are given in Chapter 4.

\chapter{Space-time of Ordinary Domain Walls}
A convenient way  to describe
the gravitational field is  in the rest frame of the wall, \ie ,
by using comoving coordinates of observers
sitting on the wall. Hence, the
wall is placed  at a fixed
$z$-coordinate, and the metric is
static in the $(t,z)$ directions transverse to the wall.
Assuming  maximal symmetry associated with the space-time internal to the
wall, the metric is taken to be  homogeneous
and isotropic  and geodesically complete
in the $(\varrho,\phi)$ surfaces parallel to the wall.
Since the extrinsic  curvature is independent of
the wall's proper time,
one can show \refmark\CGSIII\
that the metric is:
$$
ds^2=A(z)[ dt^2 - dz^2 - \beta^{-2}\cosh^2(\beta t)
d\Omega^2_{2}]\; ,
\eqn\metne
$$
with $A(z)>0$
and $d\Omega^2_{2}\equiv [1-(\beta \varrho)^2]^{-1}
d(\beta \varrho)^2+ (\beta \varrho)^2d\phi^2$.
In the extreme limit, $\beta \rightarrow 0$, the $(\varrho,\phi)$ surface
becomes a plane with $\varrho$ and $\phi$ planar polar coordinates. With
$\beta\neq 0$, the $(\varrho,\phi)$ hyperspace
is the surface of a three-dimensional
sphere, that is, its topology is ${\bf S}^2$.  In this case the coordinate
$\varrho = \beta^{-1} \sin\theta$ is compact.
The scalar curvature of the
spatial ${\bf S}^{2}$ is $2\beta^2 A(z)^{-1}[ \cosh(\beta t)]^{-2}$.
The constant $z$ section with $\beta\neq 0$
is (2+1)-dimensional de~Sitter space-time (dS$_{3}$),
which is geodesically complete.\refmark\CGSIII\
Note, that the extended nature of the wall imposes
strong constraints on the topology of the wall; the wall can be either a
planar, static configuration (Eq.\metne\ with $\beta=0$), or a time dependent
spherical bubble (Eq.\metne\ with $\beta\neq 0$).

{\bf (i) Extreme Walls}  ($\beta=0$) induce a static, conformally flat
space-time. Such walls turn out to   correspond to
supersymmetric configurations between
isolated supersymmetric minima  of  the matter potential in $4d$,
$N=1$ supergravity theory.\refmark{
\CGRI, \CG}\   Supersymmetric minima have  either zero
(Minkowski space-times ($M_4$))  or negative (anti-deSitter space-times
($AdS_4$)) cosmological constant  $\Lambda$ , which is
  related to the value of the superpotential ($W$)
and K\"ahler potential ($K$) at the minimum in the following way:
$\Lambda=-3{\kappa^{2} e^{\kappa K}|W|^2 }$.
Here, $\kappa\equiv 8\pi G$.

The existing static domain walls between
such minima have been classified:\refmark\CG\
there are two types of AdS$_4$--AdS$_4$ walls
(Type II and Type III) and an
AdS$_4$--M$_4$ wall (Type I). There are { \it no static }
(supersymmetric) domain walls between two  $M_4-M_4$ vacua.
In the following I will discuss only Type I walls and their generalizations.
This is a prototype of the walls where on one side of the wall the
space-time is asymptotically M$_{4}$; such walls
 may thus have implications for the
observable world. Generalizations to other examples, including the walls
between deSitter vacua, are discussed in  Refs.{\CGSII ,\CGSIII} .

\insrt{\hsize=2.9in
\singlespace
\vbox to 3.2in{}
\noindent
{\tenpoint{\bf Figure 1}
Penrose-Carter diagram in the $(t,z)$ direction  for the most symmetric
geodesic  extension of
the extreme $M_4-AdS_4$ domain wall configuration.
The  compact null coordinates
define the axes: $u',v' = 2\tan^{-1}[\alpha(t \mp z)]$.
These coordinates can be smoothly extended across the
nulls separating the diamonds. The domain  wall region
is denoted with the thin lines.
Cauchy horizons (dotted lines) are the nulls separating
the $AdS_{4}$  patches.}
\hrule
\vskip4mm
\normalspace
}{0}
 Extreme, thin, $M_4-AdS_4$ walls, centered
at $z=0$,  have the  energy density,  $\sigma_{{ext}}$,
and the conformal factor $A(z)$ in \metne of the following
form:\refmark{\CGRI ,\CG}
$$\eqalign{
\sigma_{{ext}}&=2\kappa^{-1}\alpha,\cr
A(z)=
 ( \alpha z-1)^{-2} &,\ \ z < 0;\ \
  A(z)=1,\ \
z > 0.}
\eqn\solI
$$
Here,  $\alpha\equiv \kappa e^{{\kappa K}\over 2}|W|=(-\Lambda/3)^{1/2}$.
The {\em horo-spherical\/}
coordinates  on the AdS$_{4}$ side are
discussed in Refs. {\CG,\Gibb} .
 The
 field equations for the matter and metric
 are coupled first order rather than second order differential equations,
thus  allowing  for a straightforward solution
for any thickness of the wall.\refmark{\CGRI, \CG}

\hangindent=\instwd\hangafter=-6
The coordinates  of the metric
\metne\  with $\beta =0$ and  $A(z)$ in \solI\
are  not geodesically complete;
geodesic extensions have been given
with emphasis on the Type I
walls in Ref.\ {\CDGS}\ and
Type II walls in Ref.\ {\Gibb}\ .
Namely,  on the $AdS_4$ side the time-like geodesics
reach  ($t = \infty ,z=-\infty$) in  a  finite
proper time $\tau = \alpha^{-1} \arcsin(1/ \epsilon) $. Here,  $\epsilon$ is
the energy per unit mass of the test particle.  Therefore,
 ($t, z$) coordinates are {\it not} geodesically complete on the $AdS_{4}$
side; there is a Cauchy horizon at $(t=\infty ,z=-\infty,)$.
The most symmetric geodesic extension (see Figure 1) comprises of
a  system of  an infinite lattice
of semi-infinite $M_4$ space-times separated by a
 $AdS_{4}$ core.
It turns out
that the $(t,r)$ line elements near the Cauchy horizon  of the
Reissner-Nordstr\"{o}m (RN) black hole and $(t,z)$ line element
 on the $AdS_4$ side of the
 wall are identical.\refmark\CDGS\
Note also a similarity between the   global
 space-time   structure (see Figure 1)  of the wall and that
of the extreme RN black hole.\Ref\HE{ See for example,
S. W. Hawking and G. F. R. Ellis, {\it The Large Scale
Structure of Space-Time}, (Cambridge University Press 1973). } However,
the time-like singularities of the RN space-time
are replaced by the domain walls.

{\bf (ii) Non-Extreme  and Ultra-Extreme Walls} induce space-time \metne\ with
$\beta \not= 0$.
In the case of thin walls  one employs  Israel's
formalism of singular hypersurfaces \Ref\ISR{
 W. Israel, Nuovo Cimento {\bf 44B},  1 (1966).}
which determines the matching conditions  across the wall region.
 Einstein's field equations and Israel's matching conditions as applied
to this case\refmark{\CGSIII}\
yield two types  of the solutions (depending on the sign of parameter $\beta$)
 with energy density
and conformal factor of the following type:
$$\eqalign{
 \sigma^{{non}}_{{ultra}}&=2\kappa^{-1}
 [(\alpha^2+\beta^2)^{1/2}+\beta],\cr
A(z)=\beta^2\alpha^{-2}
  [\sinh
  (\beta z -\beta z')]^{-2}&\ \ z<0;\ \ \
A(z)=e^{-2\beta z}\ \  z>0.}
\eqn\solII
$$
where  $e^{2\beta z'} \equiv [\alpha^2 + 2\beta^{2} +  2 \beta
 (\beta^{2} + \alpha^2 )^{1/2}]/\alpha^2 \equiv \delta$
is  determined by $A(0) \equiv 1$.

The solution with $\beta>0$   of Eq.\solII\
represents a {\em non-extreme wall.\/}
It can be shown that in the non-extreme wall region
the  potential barrier associated with the scalar field
is larger  than that for the corresponding extreme domain wall,
which implies that $\sigma_{{non}} > \sigma_{{ext}}$.
{\it E. g.,} within $N=1$ supergravity theory,
such a wall can be realized
as a wall interpolating between a
supersymmetric  M$_{4}$ vacuum and an AdS$_{4}$
vacuum with supersymmetry spontaneously broken.

In the rest frame of the wall the
non-extreme walls exhibit cosmological
horizons\refmark{\CGSII,\CGSIII}\ on both the AdS$_4$ and M$_4$ sides.
Namely, a particle  with energy per unit mass $\epsilon \ge 1$,
freely falling at constant
$\theta$ and $\phi$ in the $z \rightarrow \mp \infty$-direction
has a finite proper time\refmark\CGSII\
$ \tau =
\alpha^{-1}\{\arcsin\{ [ 1 + (\epsilon \alpha  / \beta)^{2} ]^{-1/2}
(\delta + 1)/ (\delta - 1) \}
- \arcsin [ 1 + (\epsilon \alpha  /
\beta)^{2} ]^{-1/2}\}$
and $\tau = \beta^{-1} [ \epsilon - (\epsilon^{2} - 1)^{1/2}]$, respectively.
As $\beta \rightarrow  0$, the
cosmological horizon
on the AdS$_{4}$ side becomes a Cauchy horizon (as in the extreme wall
space-time) with
$\tau = \alpha^{-1} \arcsin(1/ \epsilon) $, while the
M$_{4}$ side becomes geodesically complete.
It turns out\refmark\CGSII\ that $(t,z)$ line element on the $AdS_4$ side of
the
 wall is identical to ($t,r$) line element   near the horizon of the
non-extreme
RN black hole.

\hangindent=75.1mm \hangafter=5
In order to investigate geodesically complete
space-times for the non-extreme
walls, the metric \solII\ with is transformed\refmark\CGSII\
to  the inertial spherical
M$_{4}$ and
AdS$_{4}$ coordinates on the respective sides.
Introducing
$\underline{t} = \beta^{-1} e^{- \beta z} \sinh{\beta t}$
and
$\underline{r} = \beta^{-1} e^{- \beta z} \cosh{\beta t}$
brings the line element on the M$_{4}$ side to the spherically
symmetric form
$ds^{2} = d\underline{t}^{2} - d\underline{r}^{2} -
\underline{r}^{2} d\Omega_{2}^{2}$.
The bubble at $z=0^{+}$
lives on the hyperbolic trajectory
$\underline{r}^{2} - \underline{t}^{2} =
 = \beta^{-2}$  (see Figure 2).
On the AdS$_{4}$ side, one maps to the spherically
symmetric {\em Einstein cylinder\/} coordinates \refmark{\HE}.
This transformation is done in three steps:
(i) $\ln \Xi = \beta(z-z')$, with $0 \le \Xi \le 1$.
(ii) $T = \Xi \sinh(\beta t)$ and $R = \Xi \cosh(\beta t)$.
(iii) $T \pm R = \tan[(t_{c} \pm \psi)/2] $.
The line element on the AdS$_{4}$ side ($z<0$) becomes
$ds^{2} = (\alpha \cos \psi)^{-2}(dt_{c}^{2} -
d\psi^{2} - \sin^{2}\psi d\Omega_{2}^{2})$,
where $-\pi \le t_{c} \pm \psi  \le \pi$ and $0 \le \psi \le \pi/2$.
The bubble at $z=0^{-}$ again lives on a hyperbolic trajectory
$R^{2} - T^{2} =
= \delta^{-1}$ (see Figure 2).

\insrt{
\hsize=2.9in\singlespace
\vbox to 2.1in{}
\noindent
{\tenpoint{\bf Figure 2}
Penrose-Carter diagram for the most symmetric extension
of the non-extreme $M_4$-\break
 $AdS_4$  wall (the bubble with two insides).
 The  diagrams on the right hand side are those on the
 $M_4$ side of the bubble with  compactified coordinates
$u',v'=2\tan^{-1}[\beta(\underline{t}\mp \underline{r})]$.
The solid curved lines are the world lines of the wall bubble (its anti-podal
points). The dotted lines represent cosmological
horizons in the wall's rest frame coordinates $(z,t)$.
 The left diagram is the one of the $AdS_4$ side. It is a part of
pure $AdS_4$ cylinder\Ref\IS{See, for example, S. J. Avis, C. J. Isham and
D. Storey, Phys. Rev.
{\bf D18}, 3565  (1978).} as seen in the Einstein cylinder coordinates $(t_c,
\psi)$. The
vertical boundaries correspond to $\psi =\pi/2$ and the time direction $t_c$
is upward. The solid curved lines are
again the world lines of the wall (its anti-podal
points), sweeping one half of the fundamental domain of the pure $AdS_4$.
The dotted lines represent cosmological horizons in the wall's rest frame
coordinates  $(t,z)$.  The two types of the diagrams are identified across
the wall region.}
\hrule\vskip4mm
\normalspace}{0}
The $(t,z)$ chart is an  interpolating map which covers
the space-time on both sides of the non-extreme wall region.
To complete the space-time (see Figure 2),
one  extends on one side  onto pure M$_{4}$ .
On the AdS$_4$ side, the most symmetric  periodic extension  yields a
lattice structure of walls.  Notice that now  on
the AdS$_4$ [M$_4$] side,  the Penrose
diagram bears similarities to the one of
 a non-extreme RN
[Schwarzschild] black hole, however, {\em without\/}
singularities. The wall region of the AdS$_4$  diagram
is linked by  a wall region of the corresponding
M$_4$  diagram.
At  $t=0$  the bubble has a radius $\beta^{-1}$ which then increases as
as $\cosh(\beta t)/t$. Since the
radius of the bubble $\beta^{-1} A(z)^{1/2} \cosh(\beta t)$ decreases
as one  moves spatially away from the bubble
in both $z$ directions,  observers on {\em both\/} sides are
{\em inside\/} a bubble.

\hangindent=\instwd\hangafter=0
The  solution of  Eq.\ (\solII) with $\beta<0$
describes an {\em ultra-extreme wall.\/}
For these walls
the potential barrier associated
with the scalar field is smaller than that
of the extreme walls
which means $\sigma_{{ultra}} < \sigma_{{ext}}$ and the metric
bl\-ows up on
the M$_4$ side.
Ultra-extreme  walls exhibit the same causal structure
on the AdS$_{4}$ side as the non-extreme wall.
However, the M$_{4}$ side is geodesically complete
in the $(t,z)$-coordinates.
The M$_{4}$ side is the complement of the M$_{4}$ side.

\insrt{\hsize=2.9in
\singlespace
\tenpoint
\vbox to 2.0in{}
\noindent
{\bf Figure 3} Conformal  diagram for the classical evolution of
a false vacuum decay bubble- ultra-extre\-me  walls.
The {\bf M} region  of the M$_4$ side is covered by $(t,z)$.
The {\bf A} region is the AdS$_{4}$
discussed in Figure 2. The two diagrams are
glued at the wall region. The bubble forms at $t=t_c=0$. The jagged region
schematically indicates the quantum tunnelling process, not describable by
classical gravity.
\hrule
\vskip 4mm}{0}
The Minkowski side is on the {\em outside\/} of the
ultra-extreme bubble because the
radius\break
$|\beta|^{-1} A(z)^{1/2} \cosh(\beta t)$
increases with $z$
on the $z>0$
side. On the AdS$_{4}$ side; however,  the radius
decreases away from the wall, and thus
AdS$_{4}$ is on the
{\em inside\/} just as for the non-extreme solution.
Since $\sigma_{{ultra}} < 2\kappa^{-1}\alpha$, \REF\CGR{M. Cveti\v c, S.
Griffies and S.-J. Rey, Nucl. Phys. {\bf B389}, 3 (1993).}
i.e.\ below the Coleman-De Luccia bound,\refmark{\CD,\CGR}\
the ultra-extre\-me
solution for $t\geq 0$ describes the classical evolution of
a bubble of true AdS$_{4}$ vacuum
created by the quantum tunnelling process of false vacuum decay.
\refmark{\Coleman,\CD}
At $t=0$ the bubble is formed with
radius $|\beta|^{-1}$, expands as
$\cosh(\beta t)$, and
inevitably hits
all time-like observers on the M$_{4}$ side. The process
 is presented in Figure 3.

\chapter{Dilatonic Domain Walls}

 Dilatonic walls\refmark\CVETIC\  are pertinent to the study of $4d$
 superstring vacua with the dilaton   always arising  as an
inherent part of the supergravity fields. Thus, in the domain wall
background  (between isolated minima of the matter potential)  not only
  the metric, but also the dilaton
field may change the value.

Potentially phenomenologically viable superstring vacua
 are described by an effective  $4d$ $N=1$ supergravity theory.
The scalar part of the  effective Lagrangian  involves the  metric
$g_{E \mu \nu}$,
the  dilaton $S\equiv e^{-2\phi}+ ia$ (written in this form as
 a scalar part of the chiral superfield),  matter
matter fields and gauge fields.
 I do not include
gauge fields; however, since the dilaton  does couple to  gauge fields
 the study of  charged dilatonic walls is also interesting.
Matter fields,
 scalar components of a chiral superfield interpolate between
isolated minima of the matter potential.
\REF\Witten{E. Witten, Nucl. Phys. {\bf B268}, 373 (1986).}
\REF\WittenII{E. Witten, Phys. Lett. {\bf B155}, 151 (1985).}
\REF\FOOTII{ A superpotential for the dilaton can, however,  be induced
non-perturbatively.
 Also, the  K\" ahler potential
 receives  one-loop corrections due
to  mixed Yang-Mills-$\sigma$ model anomalies. See,
G. Cardoso and   B. Ovrut, Nucl. Phys. {\bf B369}, 351 (1992); J.P.
Derendinger, S Ferrara, C Kounnas and F. Zwirner,
Nucl. Phys. {\bf B372}, 145
(1992).
The above effects should
eventually be included in the  full treatment of
the theory.}

In the Einstein frame the effective Lagrangian of the theory can be  written in
terms of the superpotential and K\" ahler potential.  Superpotential $W_0(T)$
 is a holomorphic function of the
 matter fields $T$, only, \ie ,  to all orders in string loops  it
does not depend on the dilaton.\refmark\Witten\
In the
K\" ahler  potential $K$ the dilaton
couples\refmark{\WittenII ,\FOOTII}\ in a specific
way: $K=-\kappa^{-1}\log (S+S^*)+K_0(T,T^*)$.
The imaginary part (axion) of the dilaton  field  can be put to
zero ($a=0$);  this turns out  to be  the solution  of field
equations for the dilatonic  domain walls anyway.

A natural frame  to which strings couple is  the string frame, \ie ,
the frame of the sigma model expansion
of the string effective action. In this
case the metric in the string frame, $(g_s)_{\mu\nu}$, and the one
 in the Einstein frame, $ (g_E)_{\mu\nu}$, are related as:
 $(g_s)_{\mu\nu}=e^{2\phi}(g_E)_{\mu\nu}$. The scalar part of the Lagrangian
 is of then of the form:
$$
L_s =\sqrt{-g_s}e^{-2\phi}[ -{1 \over {2\kappa}}R_s
-2\kappa^{-1}g_s^{\mu \nu}\partial_{\mu} \phi
\partial_{\nu}\phi+ T_0 -{ V_0\over 2}]
\eqn\LS
$$
where $T_0 V_0$ correspond, respectively,  to the
  kinetic energy  and the potential
of the matter fields $T$, only.
Static planar domain walls have been found
 between isolated supersymmetric
minima of the matter potential $V_0$, whose value at the  supersymmetric
 minimum  is related to $W_0$ and
$K_0$ as: $ V_{0}\equiv
-2\kappa e^{\kappa K_0}|W_0|^2\leq 0$.
In the Einstein frame the  metric is conformally flat (see the  metric Ansatz
\metne\ with $\beta=0$) with the
conformal factor $A_E(z)$.
The  scalar field $T(z)$ and the dilaton $\phi(z)$ also depend on $z$,
only.

The minimal energy  (supersymmetric) solution  satisfies
three   first order  coupled
differential equations. Thus, straightforward solutions can
be found for any thickness of the wall
(for explicit examples see Ref.\CVETIC ).
Three types of the solutions are classified according to the value of the
potential $V_0\leq 0$ on either side of the wall.  Again,
there are {\it no} static walls where on both side of the wall
 $ V_{0}=0$.
In particular, I will describe  the  walls where on one
side of the wall $ V_{0}=0$ and on the other side of
the wall $ V_{0}<0$, \ie ,  the
 $AdS'$  -- $M'$
 walls.  $M'$ refers to the Minkowski space with $ V_0=0$ and
$AdS'$ refers to a type of anti-deSitter space-time with $ V_0<0$.

In this case the thin  wall
solution, located at $z=0$ has the explicit
form :
$$\eqalign{\sigma_{dil} & =\sqrt 2\kappa^{-1}\alpha,\cr
 A_E(z)=e^{-\sqrt2\alpha
z},&\
 \ z<0;\  \ A_E(z)=1,\  z>0. }\eqn\adsm$$
where $\alpha\equiv \kappa e^{{\kappa K_0 / 2}} |W_0 |=
(-\kappa\tilde
V_0/2)^{1/2}$.
The solution was obtained by
 normalizing the conformal factor  $A_E(0)= 1$ and choosing
the  boundary condition  $e^{2\phi(0)}=1$.
A more general  boundary condition
$e^{2\phi(0)}= e^{2\phi_0}$ allows  for a family of  one parameter
solutions.\refmark\CVETIC

 The energy density of ordinary
supersymmetric  domain walls is of a similar form (see Eq.\solI):
$\sigma_{ext} =2\kappa^{-1}\alpha$
where $\alpha\equiv \kappa e^{{\kappa K_0 \over 2}}| W_0 |
=(-\kappa
V_0/3)^{1/2}$ is defined in
terms of $W_0$ and $K_0$ in the {\it same}
 way as above.
An additional factor  $1/\sqrt 2$  in the case of dilatonic walls is
associated with the dilaton contribution
to  the quantity $ e^{{\kappa K \over 2}}| W_0 |=1/\sqrt 2 \times
e^{{\kappa K_0\over 2}}| W_0 |$. Namely,  the boundary condition
$e^{2\phi(z_0)}=1$ ensures that the effective
cosmological constant on each
side of the wall is by  a factor of $1/2$ less negative,
thus decreasing
the energy density of the wall  by a  factor of $1/\sqrt 2$. There is
a  parallel relation\refmark\HOR\ between  the mass $M$ and the charge
$Q$ for extreme RN
black holes ($M=Q$)  and extreme charged
dilatonic black holes  ($M= Q/\sqrt2$). In the domain wall  case
the role of the charge is
is played by the parameters
$\alpha$  associated with   the value of the matter
 potential at each  minimum.

\insrt{\hsize=3.0in
\singlespace
\tenpoint
\vbox to 2.4in{}
{\bf Figure 4.} Penrose diagram in the $(t,z)$ plane for
the finite size  extreme dilatonic domain wall.
The matter potential $
V_0=0$ for $z>0$  ($M'$ region) and
$ V_0<0$ for $z<0$ ($AdS'$ region).
The standard
compactified null coordinates
are defined in Figure 1.  Note the null singularity on the $AdS'$ side.
\hrule
\vskip 4mm
}{0}
 It turns out that the the solution for the conformal factor
$A_E(z)$ and the dilaton field
 imply:
$$A_s(z)\equiv A_E(z)e^{2\phi(z)}=1
\eqn\flat
$$
{\it everywhere} in the domain wall background.
\break
Therefore, the metric factor $A_s(z)$ in the string frame
 is {\it flat} everywhere.
 Although  there
is a nontrivial matter potential,
the dilaton field adjusts itself
in the domain wall background in such a way
as to leave the string  metric flat; strings do not ``feel'' the wall.

\hangindent=\instwd \hangafter=-3
The Penrose diagram   for such   walls in the $(t,z)$  plane is given
on Figure 4. The $M'$ side
corresponds to  Minkowski space-time.
 On the $AdS'$ side  (see Eq.\adsm\ )
both  the dilaton
field  and the metric  curvature blow up  as $z \to -\infty$.
 However, this  singularity
is an infinite geodesic distance away.
Note,  a formal similarity with the Penrose diagram\refmark{\CBH,\HOR}
for  the
$(r,t)$ plane of the extreme charged dilatonic black hole.

Generalizations  of the solutions to the case with
a nonpertubatively induced dilaton potential\Ref\GH{Analogous solutions
 for dilatonic black holes  with a  massive dilaton
were studied by R. Gregory and J. Harvey,
Enrico Fermi Preprint EFI-92-49 and  and
J. Horn and G. Horowitz, Santa Barbara Preprint UCSBTH-92-17, while
$2d$ black holes with a
massive dilaton were studied by  M.
McGuigan, C.Nappi, and S. Yost,
Nucl. Phys. {\bf B375}, 421 (1992). }
should also be addressed.   The case with    supersymmetry
 broken spontaneously in  the matter
part   of the potential ($ V_0$) should also be addressed.
This bears similarities to the case of
non-extreme
of charged dilatonic black holes with $M\neq Q/\sqrt 2$.
In these cases the wall need not be static anymore\refmark\CVETIC\
and the global space-time structure of such walls is of special interest.
\chapter{Conclusions}
Progress in unifying treatment of eternal  vacuum domain wall solutions
 and their  global space-time structure  was presented .
As a prototype,  the space-time of vacuum  walls between  Minkowski and
anti-deSitter space-times was addressed.
While extreme (supersymmetric) walls are planar and static configurations,
the non-extreme walls correspond to a bubble with two insides and have
energy density bounded from below by the one of the extreme wall.
Since the energy density of the extreme domain wall
is equal to the
Coleman-De~Luccia bound~\refmark{\CD}, supersymmetry provides a lower bound
\refmark\KalGib for a
(non-extreme) domain wall
separating vacua which are stable against quantum tunnelling.
On the other hand, the ultra-extreme wall,
which has energy density lower than
the one of the extreme wall, corresponds
to the classical evolution of a bubble of true
AdS$_{4}$ created by
the decay of the false M$_{4}$ vacuum.

 In addition, dilatonic walls, specific to isolated $4d$
superstring vacua, were discussed. Extreme  (supersymmetric)
 dilatonic domain walls  correspond to static configurations between isolated
supersymmetric minima of the matter potential.
 Everywhere in the domain wall background the
 dilaton field adjusts itself in a way as to
leave metric in the string frame flat.
 There are intriguing similarities between extreme dilatonic
walls and  extreme charged
dilatonic  black holes.

Presented progress in the study  of eternal domain wall configurations
provides  a step towards a
theoretical foundation for addressing
cosmological implications of domain walls, in particular, those
 arising  in supergravity and superstring
theories as supersymmetric configurations or as configurations between
isolated minima with supersymmetry spontaneously broken.

A large portion of the  work
presented here has been done in different collaborations with
S.-J. Rey, R. Davis,
and most closely  with  S. Griffies and H. H. Soleng.
The work was supported by U. S. DOE Grant No.\ DOE-EY-76-C-02-3071, and
 NATO Research Grant No. 900-700.

\refout

\end